\documentclass[11pt,a4paper]{article}
\usepackage{jheppub,amsmath,  amssymb,slashed,url,bm}
\usepackage{amsfonts}
\usepackage{graphicx}
\usepackage{epstopdf}

\title{ Renyi entropy of the critical $O(N)$ model }

\author{Anirbit}
\affiliation{ Department of Applied Mathematics and Statistics, Johns Hopkins University }
\emailAdd{amukhe14@jhu.edu}

\abstract
{ 
In this article we explore a certain definition of ``alternate quantization" for the critical $O(N)$ model. We elaborate on a prescription to evaluate the Renyi entropy of alternately quantized critical $O(N)$ model. We show that there exists new saddles of the $q$-Renyi free energy functional corresponding to putting certain combinations of the Kaluza-Klein modes into alternate quantization. This leads us to an analysis of trying to determine the true state of the theory by trying to ascertain the global minima among these saddle points. 
} 
\begin{document} 
\maketitle

\section {Definitions and folklore}  

A physical system is imagined to be separated into two parts $A$ and $B$ and let $\rho$ be the density matrix of the full system. One defines the ``reduced density matrix" of the system $A$ as $\rho_A = - Tr_B [\rho \log \rho]$.  Given a physical system $X$, $Tr_X$ is supposed to mean a trace taken in a basis of quantum states which are localized in system $X$. Then the von-Neumann entropy corresponding to $\rho_A$ is $S_A = - Tr_A [ \rho_A log \rho_A ] $ and this is defined as the ``entanglement entropy" of the system $A$. When these quantities are evaluated in any QFT then these have non-singular (believed to be ``universal") parts which are believed to be independent of the regularization prescription and are believed to contain data about a nearby CFT (``critical point").
\newpage 
But the structure of these universal terms of the entanglement entropy appear to have a strong dependence on the dimensionality of the system,

\begin{itemize}
\item For $1+1$ CFTs if the system $A$ is a system of length $x$ and $B$ is the complement of that in $\mathbb{R}$ then one can show that  $S= \frac{c}{3}\log \frac{x}{a}$ where $a$ is a short-distance cut-off and $c$ is the central charge (the universal data!) of the CFT. (here the ``=" is meant to indicate equality upto terms non-singular in the $a \rightarrow 0$ limit)  If in this $1+1$ CFT one perturbs away from the critical point then $S = {\mathcal A}\frac{c}{6}\log \frac{ \xi}{a}$ and $\mathcal{A}$ is the number of boundary points of $A$ and $\xi$ is the correlation length. (one is imagining the $A$ and the $B$ to be composed of intervals each of whose lengths is larger than $\xi$) 

\item For CFTs in dimensions $1+(d>1)$ it is conjectured that the leading contribution to the entanglement entropy $S$ takes the form of what is called the ``area law". This means that the most divergent piece of $S$ is $f(a)\frac{\mathcal{A} }{ a^{d-1}}$ where $\mathcal{A}$ is the area of the boundary of the region $A$ and and $f$ is some function of the short-distance cut-off $a$. This believed ``area law" on one hand means that the entanglement entropy is localized on the boundary of $A$ and on the other hand it also means that there is nothing universal about this leading (most divergent) piece of $S$. One believes that at least in the case of $1+2$ CFTs when the boundary between $A$ and $B$ is a closed smooth manifold the difference (say $\Delta S$) $= S -  f(a)\frac{\mathcal{A} }{ a^{d-1}}$ is something universal (and hopefully geometric). Also the expectation is that this $\Delta S$ remains universal even when the theory is moved away from the critical point. In the case of critical $O(N)$ model  in $2+1$ dimensions this universality is seen in both $\frac{1}{N}$ and $\epsilon = 3-d$ expansions and the universality of this $\Delta S$ is expected to extend into the range $1< d <4$. \cite{Max} 

\item On deforming away from the critical point it is conjectured that the this entanglement entropy should change as,  $lim _{L \rightarrow \infty} S = f(a) \left (  \frac{L}{a} \right ) ^{d-1} + r \left ( \frac{L}{\xi} \right )^{d-1}$ where $\xi$ is the correlation length and $L$ is some length scale of the entangling surface.  

\item These conjectures about the universal parts also extend to the Renyi entropies defined as $S_q = \frac{1}{1-q} \log \left [ Tr_A[ \rho_A^q] \right ]$ (and then $S = lim _{q \rightarrow 1} S_q$). Now one expects that on deformation from a CFT the correlation length is seen by the $S_q$ as, $lim_{L \rightarrow \infty} S_q = f_q(a) \left (  \frac{L}{a} \right )^{d-1} + r_q \left (  \frac{L}{\xi}  \right ) ^{d-1}$(the coefficients $f$ and $r$ now have a $q$ dependence) 
\end{itemize} 

\subsection {Holography improved conjectures for spherical entangling surfaces}  

Just QFT considerations go only so far but with the help of AdS/CFT \cite{M},  Ryu-Takyanagi proposed a more detailed structure for the universal parts of the entanglement entropy \cite{R1,R2,R3}. Let $A$ be a $d-1$ dimensional spatial entangling surface in the $d+1$ dimensional space-time at the boundary of the bulk space-time.  The CFT of interest lives on this boundary and the bulk is asymptotically locally $AdS_{d+2}$. Let $\gamma_A$ be the minimal surface in the bulk which has $A$ as its boundary. Then the Ryu-Takyanagi conjecture states that, $S_A = \frac{ Area (\gamma_A)}{4G_N^{d+2}}$. This conjecture obviously reminds one of the Bekenstein-Hawking entropy formula and the two things indeed seem related in various ways. \cite{BHEE1,BHEE2,BHEE3,BHEE4}\\
~\\
It needs to be emphasized that the Ryu-Takyanagi conjecture uses only a tiny amount of the full power of Maldacena's duality. In particular in scenarios of $AdS_3/CFT_2$ where it has been extensively checked one needs to take the limit of large central charge on the CFT side and by virtue of the Brown-Henneux formula we understand that to be a ``semi-classical" limit. It is an important question to understand the possible generalization of the Ryu-Takyanagi conjecture at finite central charge.\cite{CC1,CC2,CC3,CC4,CC5} \\
~\\
Using the Ryu-Takyanagi conjecture in the case of a spherical entangling surface one believes that the divergent and the universal parts of the entanglement entropy for a CFT in odd $D$ space-time dimensions is of the form, $S =  s_{D-2} \left (  \frac {R}{\epsilon} \right ) ^{D-2} + ... + s_1 \frac{R}{\epsilon} + s_0 + O(\epsilon)$ and for $D$ even, $S = s_{D-2} \left ( \frac {R}{\epsilon} \right ) ^{D-2} + ...+ s_2 \left ( \frac{r}{\epsilon} \right ) ^2 + s_L ln \left ( \frac{2R}{\epsilon} \right ) + \tilde{s_0} + O(\epsilon^2)$. The $s_0$ and the $s_L$ are the expected universal terms. $s_0$ is proportional to the partition function of the Euclidean CFT on $S^D$. $s_L$ is proportional to the central charge defined from the coefficient of the Euler density in the trace anomaly.

\subsection {Mapping to thermal CFTs on hyperbolic cylinders}  

An argument of Casini, Huerta and Myers \cite{CHM}, shows that the reduced density matrix $\rho_A$ for the special case of a spherical entangling surface of radius $R$ is the same as the thermal partition function of the $CFT_{d}$ on a hyperbolic cylinder ($\mathbb{H}_{d-1} \times \mathbb{R}$) of radius $R$ at a temperature $T = \frac{1}{2\pi R}$. Then the entanglement entropy becomes the thermal entropy of the Eulcidean $CFT_d$ on $\mathbb{H}_{d-1} \times S^1$ where the circle is of radius $\frac{1}{2\pi R}$ and the $\mathbb{H}_{d-1}$ continues to be of radius $R$. By an extension of the same argument Smolin, Myers, Jung and Yale \cite{HMSY} have shown that similarly Renyi entropies can also be written as, $S_q = \frac{2\pi qR}{1-q} ( F( \frac{1}{2\pi R} ) - F(\frac{1}{2\pi qR})  )$, where $F(T)$ is the free energy of the $CFT_{d}$ on the same $\mathbb{H}_{d-1} \times \mathbb{R}$ of radius $R$ at a temperature $T$.

\section { Effect of double-trace deformation in holographic CFTs  } 

If one considers a scalar field on $AdS_{d+1}$ (we will assume it to be unit radius unless stated otherwise) then one can show that if the field decays towards the boundary as $z^\Delta$ ($z$ as defined in the usual Poincare patch coordinates) then the only way the field can be normalizable is if $\Delta > \frac{d}{2}$. But one knows that the unitarity bound of a $CFT_d$ is $\frac{(d-2)}{2}$ and by $AdS/CFT$ this $\Delta$ is supposed to be the dimension of the boundary operator dual to this bulk scalar field. So naively there seems to be a gap between these two facts.\\
~\\
So one asks if it is possible to somehow change the bulk action for the scalar field to get the normalizable bulk solutions to asymptote as powers less the $\frac{d}{2}$. This is achieved by the Klebanov-Witten form \cite{KW} of the scalar action gotten by dropping the total derivative from the bulk scalar action.   (modulo the subtleties with the Gibbons-Hawking term if the spacetime has a boundary) Then the new achieved lower bound on the asymptotic powers of the normalizable modes matches the $CFT_d$ unitarity bound of $\frac{d-2}{2}$
\newpage 
But a more curious thing happens once this new shifted Klebanov-Witten action is used in the bulk. Initially the two possible boundary asymptotics for the bulk scalar field were $\Delta_{\pm} = \frac{1}{2} (d \pm \sqrt{d^2 + 4m^2})$ and only one of them was above the normalizablity enforced lower bound on the bulk asymptotics. But once this lower bound falls to $\frac{d-2}{2}$ one sees that there now appears a mass range $(-\frac{d^2}{4}, -\frac{d^2}{4} + 1)$ where both the possible solutions for $\Delta$ are above the lower bound. ($-\frac{d^2}{4}$ is what is called the Breitenlohner-Freedman bound)\\
~\\
So if one is sitting in this sweet spot of mass range where both the solutions are above the unitarity (boundary)/normalizability (bulk) bound one is tempted to ask if there is a continuous deformation on the boundary which RG flows between one fixed point to the other.  And that exists and they are given by ``double-trace" deformations of the form  $O^2$ where $O$ is an operator of dimension in the range $(\frac{d}{2}-1, \frac{d}{2})$. This is a ``relevant" deformation being done in the UV so that the it affects the IR without affecting the UV. The deformation strength needs to vanish in the UV and go to infinity at the IR where it will hit the fixed point.\\
~\\
Klebanov and Witten go on to show that these two CFTs (when both exist!) have generating functionals related by Legendre transforms. When the bulk scalar goes as $z^{\Delta_{+}}$ it is detecting the IR fixed point on the boundary and when it goes as $z^{\Delta_{-}}$ it is detecting the UV fixed point on the boundary. So $\Delta_{+}$ is the conformal dimension of the dual operator in the IR and $\Delta_{-}$ in the UV.  (one has $\Delta_+ + \Delta_- = d$)\\
~\\
One can expand this discussion beyond just scalar fields in the bulk and consider spin-s (Bosonic) fields in the bulk. In the following we briefly sketch the general ideas following the beautiful paper, \cite{SS} For $s \geq 1$ the unitarity bound in $AdS_{d+1}$ will be $\Delta = s + d -2$ and the two possible asymptotics in small-$z$ for the (free) spin-s fields are $z^{-s + ( \Delta_{-} = 2 - s)}$ and $z^{ -s + (\Delta_{+} =  s + d -2)}$  (these are quite distinct from the $s=0$ case described earlier!). So when the bulk spin-s field is asymptoting as $(-s +  \Delta_{-})$, clearly the dual operator in the boundary CFT (sitting at its UV fixed point) has to have conformal dimension of $\Delta_{-}$ and the most natural such choices are the spin-s gauge-fields.  And similarly at the IR fixed point the dual operator will have a conformal dimension of $\Delta_{+}$, which is the unitarity bound and the dual operator will be a spin-s conserved current.\\ 
~\\
One wants to understand if the two scenarios of having boundary conserved currents in the IR and boundary spin-s gauge fields in the UV can be connected by a double-trace deformation induced RG flow. One thinks of deforming the boundary theory in the IR by a double-trace term of the form, $\int (J^{(s)})^2$, where $J^{(s)}$ is a spin-s operator of dimension $\Delta  > \frac{d}{2}$. So its an irrelevant deformation being done in the boundary in the IR. This will induce a RG flow which will gives rise to an UV fixed point at which the dimension of $J^{(s)}$ will be $\Delta_{-} = d - \Delta + O(1/N)$. (thinking of a boundary theory made up of $N$ complex scalar fields on $S^d$) So now specializing to the case $d=3$ and recalling the unitarity bound for $s \geq 1$ as stated earlier one sees that for the UV theory to be unitary one needs, $3 - \Delta \geq s+1$. Combining  this with our condition for the deformation to be irrelevant we have, $ \frac{3}{2} < \Delta \leq 2-s$.  Its clear that this condition can't be satisfied for $s \geq 1$.\\
~\\
One special case of the above scenario is when $d =3$ and $\Delta = s+1$ and then $J^{(s)}$ in the UV becomes a spin-s gauge field of dimension $2-s$ and then its not a gauge singlet and hence not an observable and hence doesn't raise any obvious reason to comply with the unitarity bounds. Then in the UV we have a conformal spin-s gauge field theory. Another special case is when $s=0$ and then the unitarity bound is $\frac{d-2}{2}$ and hence for the UV to be unitary one needs $3 - \Delta > \frac{3-2}{2}$ and hence $\Delta > \frac{5}{2}$.\\
~\\
One wants to understand the change in the free-energy between the UV and the IR fixed points under such a double-trace transformation and the argument clearly needs to be different depending on whether $\Delta \neq s+1$ or $\Delta = s +1$ (restricting to $d=3$). Then reproduced from both the bulk as well as from the boundary in \cite{SS} it has been argued that the free-energy change is given by, 

\begin{itemize}
\item When $\Delta \neq s+1$, 
$$ F^{(s)}_{UV} - F^{(s)}_{IR} = -\frac {(2s+1)\pi}{6} \int _{\frac{3}{2}}^{\Delta} dx (x-\frac{3}{2}) (x+s-1)(x-s-2)Cot(\pi x) $$

\item When $\Delta = s+1$, 
$$F^{(s)}_{gauged\text{ }UV} - F^{(s)}_{free\text{ }IR} = \frac{ (4s^2-1)s }{6}log N + O(N^0) $$
\end{itemize}

\section { Renyi entropy in large-$N$ (critical) spherical non-linear sigma model }

Let us focus on the $(\phi^2)^2$ field theory. We are thinking of this as the free scalar field theory being deformed in the UV by a double-trace term of the form $\delta S^{UV} \sim \int d^3x J^{(0)}J^{(0)} $ where $J^{(0)} = \sum_{i=1} ^N \phi^i \phi^i$ is a scalar singlet operator of scaling dimension $\Delta = 1$ in the UV.  It is known that for a double-trace deformation like this in the UV to flow to to a  IR fixed point one needs the operator to have UV dimensions in the range $(\frac{d}{2}-1, \frac{d}{2})$ and here one sees that this will satisfy the condition for $d=3$. For such a deformation it has been shown for the free energy ($F = - \log Z$) both from the bulk (a Vasiliev theory on $AdS_4$) as well as from the boundary that $\delta F =  F^{UV} -  F^{IR} = - \frac{\xi(3) }{8\pi^2} + O(\frac{1}{N}) $ \\
~\\
We now try to go further than this and try to understand the Renyi entropies at the large-$N$ critical point of this theory.  This system is potentially interesting because here at the large-$N$ critical point one can mainatin a dual gravity description and also get quantum corrections in the boundary. This can be believed because the large-$N$ limit of this $\phi^4$ theory is exactly equal to the large-$N$ spherical non-linear sigma model to all orders in $\frac{1}{N}$ and hence if the former is critical so is the later. And we have that for critical non-linear sigma model the central charge is proportional to the dimension of the target manifold which in this case is $S^{N-1}$.\newpage
Hence the large-$N$ critical theory has its central charge increasing (linearly with $N$) and hence one can believe that a consistent gravity truncation should holographically exist in the bulk.\\
~\\
The recent papers by Maldacena, Aitor, Faulkner \cite{CC6} and Hartman \cite{CC2} have made this point amply clear that Ryu-Takyanagi conjecture is a (semi)classical phenomenon in the (boundary) bulk. Ryu-Takyanagi is a shadow evidence that there \emph{might} be a holographic picture out there. That the conjecture is classical in the bulk is obvious right away since nothing more than classical gravity is used there. On the boundary side the thing is semi-classical in a way that has been made clear in Hartman's recent paper - that one needs to take a double limit whereby the primary operator dimension and the central charge are both very large but at a constant ratio. And large central charge is basically suppressing the quantum effects - as can be evidenced from say Brown-Henneux like arguments. \\
~\\
We write the partition function of the non-linear sigma model as, 

\begin{eqnarray} \label{nlsm} 
Z_q =  \int [D\phi(x)][D\lambda(x)] \exp [ - S(\phi, \lambda)  ]
\end{eqnarray}

where 

\begin{eqnarray}
S(\phi, \lambda) = \frac{1}{2g} \int_{ \mathbb{H}^2 \times S^1_q  }  d^d x \sqrt { g}  [ (\partial_\mu \phi)^2 + \lambda (x)  (\phi^2 - N)]
\end{eqnarray} 

and $\phi$ is a $N$ component vector field and $\lambda$ is a Lagrange multiplier which is constraining the field to be on a $S^{N-1}$ defined as $\phi^2 = N$. We are using the imagery of a $\phi$ representing a classical spin of size $\sqrt{N}$ and the coupling $g$ is the loop expansion parameter. As necessary for the calculation of the $q-$Renyi entropy the action integral is being done on $\mathbb{H}^2 \times S^1_q$ (where $\mathbb{H}^2$ is the $2-$dimensional hyperbolic plane or Euclidean $AdS_2$ and $S^1_q$ is a circle with the standard round metric on it but where the angular parameter goes from $0$ to $2\pi q$)\\
~\\
One might want to make it explicit that the Lagrange multiplier field of $\lambda(x)$ is implementing a point-wise constraint on the base manifold as in, $\int D \lambda \exp [-\frac{1}{2g} ] \int d^dx \sqrt{g} \lambda(x) ( \phi^2 (\vec{x}) - N  ) \sim \prod_{x} \delta (\phi^2 ( \vec{x} ) - N  )$\\
~\\
Hence doing the Gaussian integral over $\phi$ we have, 

\begin{eqnarray} 
Z_q = \int D\lambda \exp [-\frac{N}{2} Tr \log [ \frac{ - \partial ^2 + \lambda  }{g}  ]   ] \exp [ \frac{N}{2g} \int d^dx \sqrt{g} \lambda   ] 
\end{eqnarray}

Now we look for uniform saddles of the kind, $\lambda (x) = m^2 - \frac{1}{4} $. Here we think of $m^2$ as the conformal mass on the branched manifold and we are measuring it in terms of its gap from the BF bound on $\mathbb{H}^2$ of unit radius i.e $-\frac{1}{4}$. 
\newpage 
So at the large-N saddle we have, 

\begin{eqnarray} 
Z_q = \exp [-\frac{N}{2} Tr \log [ \frac{ - \partial ^2 + m^2 - \frac{1}{4}  }{g_c}]  +  \frac{N}{2g_c} Vol ( \mathbb{H}^2 \times S^1_q )(m^2 - \frac{1}{4} )   ] 
\end{eqnarray}

We note that from here on we make it explicit that the coupling constant $g$ is tuned to its flat-space critical value of $g_c$ and that its only the large-N theory that is critical and its the critical theory which has been put on the branched manifold. The main goal of the following analysis will be to see if and there can exist such a conformal mass for this flat-space critical theory lifted to the branched manifold. \\
~\\
Now we have for the free-energy $F_q$ as, 

\begin{eqnarray} 
F_q = \frac{N}{2}Tr_{ \mathbb{H}^2 \times S^1_q} \log [ -\partial^2 + m^2 - \frac{1}{4} ] - \frac{N}{2g_c}  Vol ( \mathbb{H}^2 \times S^1_q )(m^2 - \frac{1}{4} ) 
\end{eqnarray} 

For a circle factor the eigenfunctions are $e^{\frac{ inz } {q}} $ (thinking of $z$ as the coordinate around the circle). The volume of this circle is $2\pi q$. We continue with the conventions from \cite{SS} about the measure and eigenvalues of the scalar harmonics on $\mathbb{H}^2$ and hence we have, 

\begin{eqnarray} 
\frac { 2F_q }{ N Vol( \mathbb{H}^2 ) } = \sum_{ n \in \mathbb{Z} } \int_{\mu >0}  d\mu \frac{ \tanh (\pi \sqrt{\mu} ) }{4 \pi  } \log ( \mu + m^2 + \frac{n^2}{q^2}  ) - \frac{2\pi q }{g_c}  (m^2 - \frac{1}{4}  )
\end{eqnarray}

The conformal mass ($m^2$) of this theory is that value of $m^2$ for which $F_q$ extremizes i.e the value of $m^2$ for which the following derivative vanishes, 

\begin{eqnarray} 
\frac{2 } {N Vol (\mathbb{H}^2 )  } \frac {\partial F_q  } {\partial m^2  }  =  \sum _ { n \in \mathbb{Z} } \frac{1}{4\pi} \int_{\mu >0} d\mu \frac{\tanh (\pi \sqrt{\mu}  )  }{ \mu + m^2 + \frac{n^2}{q^2}  }  - \frac{2 \pi q } {g_c}
\end{eqnarray} 

But this infinite sum is term by term divergent! So we need to come up with a regularization scheme for this and that is to $(1)$ first sum over the infinite KK modes and then $(2)$ to take the saddle-point of the difference free-energy, $F_q(m^2) - qF_1 (0)$. The intuition being that at $q=1$ the theory is on flat-space where the conformal mass vanishes and the physically relevant part of the free-energy of the critical theory on the $q-$branched manifold is its difference from $q$ copies of the flat-space free-energy of the ``same" critical theory. \\
~\\
For some $a, q \in \mathbb{R}$ we have the zeta-function regularized identity,
\begin{eqnarray*}
\sum_{n \in \mathbb{Z} } \frac{ 2 a}{ a^2 + \frac{n^2}{q^2}  } = 2 \pi q \text{ }sgn (a) \coth ( \pi q \vert a \vert  )
\end{eqnarray*}
Now we use this in the above saddle equation to get, 

\begin{eqnarray}
\frac{2}{N Vol(\mathbb{H}^2)}\frac{ \partial F_q} {\partial m^2  } = \frac{q}{4} \int_{ \mu >0}  d\mu \tanh ( \pi \sqrt{\mu}  )\text{ }sgn (\sqrt{\mu + m^2 }  )  \frac { \coth ( \pi q \vert  \sqrt{\mu + m^2 }  \vert  )   } { \sqrt{\mu + m^2 }   }  -  \frac{2 \pi q}{g_c} 
\end{eqnarray} 

These integrals are still divergent and we now focus on the physically motivated difference. (and choosing the positive square-root) we have, 

\begin{eqnarray}
\frac{2}{N Vol(\mathbb{H}^2)} \left [    \frac{ \partial F_q} {\partial m^2  }   - q (   \frac{ \partial F_q} {\partial m^2  }  \vert _{ q=1,m=0}   )  \right ] \\ 
\notag = \frac{q}{4} \int_{ \mu >0}  d\mu \tanh ( \pi \sqrt{\mu}  )\left [   \frac { \coth ( \pi q \sqrt{\mu + m^2 }    )   } { \sqrt{\mu + m^2 }   }  -  \frac { \coth ( \pi \sqrt{\mu }  )   } { \sqrt{\mu }   } \right ] 
\end{eqnarray} 

The above quantity is finite! One can flow the R.H.S above as a function of $m$ and see that $(1)$ it monotonically decreases with increasing $m>0$ as well as for decreasing $m<0$ (it maximizes for $m=0$) and $(2)$ that it has a root in $m$ only for $q\leq 1$ (at $q=1$ its the unique root $m=0$ whereas for $q<0$ there are two roots symmetrically on the positive and the negative $x-$axis). For all $q>1$ the function is forever negative. \\
~\\
So we see that doing the ``standard" large-N $O(N)$ model when put on the branched manifold doesn't have a critical point and hence its not a CFT there. So we need to do some modification to this theory to get critical theories on the branched manifold. 

\subsection {Alternate quantization for a Kaluza-Klein mode} 

Let us take a closer look at the contribution of each KK mode to the above analysis. We remember that to the quantity, $\frac{2}{N Vol( \mathbb{H}^2 ) } \frac { \partial F_q (m^2) } {\partial m^2 }$ the $n^{th}$ KK mode contributed the (divergent) quantity, 

\begin{eqnarray}
\frac{1}{4 \pi } \int _ {\mu >0 } d\mu \frac { \tanh ( \pi \sqrt{\mu} )  } { \mu + m^2 + \frac {n^2}{q^2}  }  =  \frac{1}{2 \pi} \int _ {\lambda >0} d\lambda \frac {\lambda \tanh ( \pi \lambda )  } {  \lambda ^2 + m^2 + \frac{ n^2}{q^2} } 
\end{eqnarray}

Firstly we imagine this to be an integral analytically continued to the complex $\lambda$ plane. Now we regulate the integral by putting an UV cut-off $\Lambda$ on the eigen-spectrum. Then in this complex picture using the parity symmetry in $\lambda$ we see that the value of this integral is ($2\pi i$ times) half the sum of the residues of the integrand in a semicircle of radius $\Lambda$ about the origin and the semi-circle being closed either in the upper or the lower half plane. One notes that on that plane the function $\frac {\lambda \tanh ( \pi \lambda )  } {  \lambda ^2 + m^2 + \frac{ n^2}{q^2} }$ has poles at $i(n + \frac{1}{2})$ ($n \in \mathbb{Z}$) and at $\pm \vert  \sqrt{ m^2 + \frac{n^2}{q^2}  }  \vert$ and the residues of it at these poles are respectively,  $\frac{-2i (2n +1 ) } {\pi ( 1 + 4( n^2 + n - ( m^2 + \frac{n^2}{q^2}  ) ) )   } $ and $\pm \frac{i}{2} \tan \left [ \vert  \sqrt { m^2 + \frac{n^2}{q^2}  }  \vert \pi  \right ]$.  \\
~\\
In the $AdS_{d+1}$ what constituted as the choice of standard vs alternate boundary conditions was the choice between the boundary scaling of the fields to be either $z^{\Delta_\pm}$. In the context of this integral, we say that the difference between the two quantization is to be thought of as the choice between which of the ``mass" poles ( $\pm \vert  \sqrt{ m^2 + \frac{n^2}{q^2}  }  \vert$  )  is to be accounted for while keeping the semi-circular contour in the upper half-plane.    \\
~\\
So if we are are going to put the $n^{th}$ KK mode in alternate quantization then from the quantity previously calculated, $\frac{2}{N Vol(\mathbb{H}^2)} \left [    \frac{ \partial F_q} {\partial m^2  }   - q (   \frac{ \partial F_q} {\partial m^2  }  \vert _{ q=1,m=0}   )  \right ]$ we first subtract the contribution of the $n^{th}$ KK mode in the standard quantization and then add in the contribution in the alternate quantization.  This expression is given as, 

\begin{eqnarray}
\notag &\frac{q}{4} \int_{ \mu >0}  d\mu \tanh ( \pi \sqrt{\mu}  )\left [   \frac { \coth ( \pi q \sqrt{\mu + m^2 }    )   } { \sqrt{\mu + m^2 }   }  -  \frac { \coth ( \pi  \sqrt{\mu }  )   } { \sqrt{\mu }   } \right ]\\
&+ \frac{1}{2\pi} \left [     -   \int _ {\lambda >0, in}^{ \Lambda \rightarrow \infty   }  d\lambda \frac {\lambda \tanh ( \pi \lambda )  } {  \lambda ^2 + m^2 + \frac{ n^2}{q^2} }       +      \int _ {\lambda >0, out}^{ \Lambda \rightarrow \infty   }  d\lambda \frac {\lambda \tanh ( \pi \lambda )  } {  \lambda ^2 + m^2 + \frac{ n^2}{q^2} }    \right ] 
\end{eqnarray}  

In the above expression the subscript $in/out$ means whether when interpreted as a complex integral (along a semi-circular contour in the upper-half plane) which of the poles $\pm \vert  \sqrt{ m^2 + \frac{n^2}{q^2}  }  \vert$ is taken to contribute respectively. One notes that each of the integrals without the UV cut-off of $\Lambda$ are divergent but we realize that a net finite contribution is obtained by first taking at the difference between the integrals with a cut-off and them later letting the cut-off go to infinity. \\
~\\

So we have, 

\begin{eqnarray} 
\notag &\int _ {\lambda >0, in/out}^{ \Lambda  }  d\lambda \frac {\lambda \tanh ( \pi \lambda )  } {  \lambda ^2 + m^2 + \frac{ n^2}{q^2} }\\
\notag &= \frac{2 \pi i}{2} \left [   \sum _{n=0}^{\Lambda}  \frac{-2i (2n +1 ) } {\pi ( 1 + 4( n^2 + n - ( m^2 + \frac{n^2}{q^2}  ) ) )   }    \pm \frac{i}{2} tan ( \vert  \sqrt { m^2 + \frac{n^2}{q^2}  }  \vert  \pi  )   \right ] \\
\notag &= \frac{2 \pi i}{2} \left [    \frac{i}{2\pi}  \left [   H_{-\frac{1}{2} -  \vert  \sqrt { m^2 + \frac{n^2}{q^2}  }  \vert } +  H_{-\frac{1}{2} +  \vert  \sqrt { m^2 + \frac{n^2}{q^2}  }  \vert } - H_{\frac{1}{2} + \Lambda -  \vert  \sqrt { m^2 + \frac{n^2}{q^2}  }  \vert  } - H_{ \frac{1}{2} + \Lambda +  \vert  \sqrt { m^2 + \frac{n^2}{q^2}  }  \vert  }           \right ] \right ]  \\
&+\frac{2 \pi i}{2} \left [  \pm \frac{i}{2} tan ( \vert  \sqrt { m^2 + \frac{n^2}{q^2}  }  \vert  \pi  )    \right ] 
\end{eqnarray} 

In the above the $+/-$ on the RHS corresponds to the $in/out$ on the LHS respectively. And the $H$ stand for the harmonic number function.  Now doing a large $\Lambda$ asymptotic expansion of R.H.S about $\Lambda = \infty$ we have, 

\begin{eqnarray} 
\notag \int _ {\mu >0, in/out}^{ \Lambda \rightarrow \infty  }  d\lambda \frac {\lambda \tanh ( \pi \lambda )  } {  \lambda ^2 + m^2 + \frac{ n^2}{q^2} } &= \frac{2 \pi i}{2} \left [  \frac{i}{2 \pi} (  -2 \gamma + H_{-\frac{1}{2} - \vert  \sqrt { m^2 + \frac{n^2}{q^2}  }  \vert   } + H_{-\frac{1}{2} + \vert  \sqrt { m^2 + \frac{n^2}{q^2}  }  \vert   }   ) - \frac{i}{\pi}log \Lambda + O(\frac{1}{\Lambda})  \right ] \\
 &  +\frac{2\pi i } { 2} \left [ \pm \frac{i}{2} \tan ( \vert  \sqrt { m^2 + \frac{n^2}{q^2}  }  \vert  \pi  )     \right ] 
\end{eqnarray} 

So we see that when the ``out" integral is subtracted from the ``in" integral the UV divergence parameterized by $\Lambda$ will cancel out and we will be left with a regular contribution. Hence we can write the regularized saddle equation for the $n^{th}$ KK mode being in alternate quantization as, 

\begin{eqnarray}
\notag &\frac{2}{N Vol(\mathbb{H}^2)} \left [    \frac{ \partial F_q} {\partial m^2  }   - q (   \frac{ \partial F_q} {\partial m^2  }  \vert _{ q=1,m=0}   )  \right ]^{(n,alternate)} = \frac{q}{4} \int_{ \mu >0}  d\mu \tanh ( \pi \sqrt{\mu}  )\left [   \frac { \coth ( \pi q \sqrt{\mu + m^2 }    )   } { \sqrt{\mu + m^2 }   }  -  \frac { \coth ( \pi q \sqrt{\mu }  )   } { \sqrt{\mu }   } \right ]\\
\notag &+ \frac{i}{2} \left [     -     \left \{     \frac{i}{2 \pi} (  -2 \gamma + H_{-\frac{1}{2} - \vert  \sqrt { m^2 + \frac{n^2}{q^2}  }  \vert   } + H_{-\frac{1}{2} + \vert  \sqrt { m^2 + \frac{n^2}{q^2}  }  \vert   }   ) - \frac{i}{\pi}log \Lambda + O(\frac{1}{\Lambda})      + \frac{i}{2} tan ( \vert  \sqrt { m^2 + \frac{n^2}{q^2}  }  \vert  \pi  )      \right \}    \right ] \\
\notag &+  \frac{i}{2} \left [    \left \{   \frac{i}{2 \pi} (  -2 \gamma + H_{-\frac{1}{2} - \vert  \sqrt { m^2 + \frac{n^2}{q^2}  }  \vert   } + H_{-\frac{1}{2} + \vert  \sqrt { m^2 + \frac{n^2}{q^2}  }  \vert   }   ) - \frac{i}{\pi}log \Lambda + O(\frac{1}{\Lambda})   -   \frac{i}{2} tan ( \vert  \sqrt { m^2 + \frac{n^2}{q^2}  }  \vert  \pi  )     \right \}  \right ] 
\end{eqnarray}  

and this simplifes to, 

\begin{eqnarray}
\notag &\frac{2}{N Vol(\mathbb{H}^2)} \left [    \frac{ \partial F_q} {\partial m^2  }   - q (   \frac{ \partial F_q} {\partial m^2  }  \vert _{ q=1,m=0}   )  \right ]^{ (n,alternate)}\\
\notag & = \frac{q}{4} \int_{ \mu >0}  d\mu \tanh ( \pi \sqrt{\mu}  )\left [   \frac { \coth ( \pi q \sqrt{\mu + m^2 }    )   } { \sqrt{\mu + m^2 }   }  -  \frac { \coth ( \pi \sqrt{\mu }  )   } { \sqrt{\mu }   } \right ] + \frac{1}{2}\tan ( \vert  \sqrt{ m^2 + \frac{n^2}{q^2}   }   \vert \pi ) 
\end{eqnarray} 

\subsection {All combinations of putting a KK mode in alternate quantization} 

We note that given the effective notion of mass for the $n^{th}$ KK mode its BF bound is $-\frac{n^2}{q^2}$ (so the standard quantization for it is valid in the range $-\frac{n^2}{q^2} \leq m^2$). Further it can be put in alternate quantization for $-\frac{n^2}{q^2} \leq m^2 \leq 1 - \frac{n^2}{q^2}$.  So the $n=0$ mode has the highest BF bound of $0$ and hence the $n>q$ modes can never be put in the alternate quantization since that will have no common mass range to share with the other modes. Further if the $n=q$ mode is put in alternate then its upper bound will coincide with the lower bound of the $n=0$ mode and hence irrespective of the quantization of the other modes only the $m^2 = 0$ situation will be allowed. Hence modulo these considerations, the modes $n=0,...,q-1$ can be assigned either the standard or the alternate quantizations and hence we have $2^{q}$ possible scenarios to explore at every $q$. \\
~\\
Hence we write the expression for the saddle point as, 

\begin{eqnarray}\label{saddle}
&\frac{2}{N Vol(\mathbb{H}^2)} \left [    \frac{ \partial F_q} {\partial m^2  }   - q (   \frac{ \partial F_q} {\partial m^2  }  \vert _{ q=1,m=0}   )  \right ]\\
\notag & = \frac{q}{4} \int_{ \mu >0}  d\mu \tanh ( \pi \sqrt{\mu}  )\left [   \frac { \coth ( \pi q \sqrt{\mu + m^2 }    )   } { \sqrt{\mu + m^2 }   }  -  \frac { \coth ( \pi \sqrt{\mu }  )   } { \sqrt{\mu }   } \right ] + \sum _{n=0} ^{n=q} s_n \frac{1}{2}\tan ( \vert  \sqrt{ m^2 + \frac{n^2}{q^2}   }   \vert \pi ) 
\end{eqnarray} 

The parameters $s_n$ are put in such that  if $s_n =1$ then it would put the $n^{th}$ KK mode in the alternate quantization and if $s_n =0$ then it would have that mode in the standard quantization. In the above we have let the sum on the RHS to go till the $q^{th}$ mode with the implicit understanding that is $s_q =1$ then the only value of $m^2$ that can be tested as a candidate saddle is $m^2=0$. 
\newpage 
We remember that if $s_j = 0$ then its allowed range is $[-\frac{j^2}{q^2},\infty)$ and if $s_j = 1$ then its allowed range is $[-\frac{j^2}{q^2},1-\frac{j^2}{q^2}]$. Thus the valid range of $m^2$ is the intersection of all these constraints (one constraint each for each of $s_{n=0,..,q}$).\\
~\\
It turns out that this integral in the above expression is well-defined and can be numerically evaluated for specific values of the parameters. Care needs to be taken that the integrand be supplied to the computer as $\frac{q}{4} \int_{ \mu >0}  d\mu \tanh ( \pi \sqrt{\mu}  )\left [   \frac { \sqrt{\mu} \coth ( \pi q \sqrt{\mu + m^2 })  -  \sqrt{\mu + m^2 }\coth ( \pi \sqrt{\mu }  ) } { \sqrt{\mu} \sqrt{\mu + m^2}  } \right ]$. Hence one can numerically find the roots of \ref{saddle}.\\ 
~\\
For example at $q=2$  the allowed ranges and saddles are, \\
~\\

\label{q2} 
\begin{tabular} { | c | c | c | c | c | } 
\hline
n = 0  &   n= 1 & n = 2 & allowed $m^2$ range & saddle value of $m^2$  \\ 
\hline 
S & S & S & $[0,\infty)$ &  no solutions \\
\hline
A & S & S & $[0,1]$ & $\sim 0.027$ \\
\hline
S & A & S & $[0,\frac{3}{4}]$ & no solutions \\
\hline
A & A & S & $[0,\frac{3}{4}]$ &  $\sim 0.160$  \\
\hline
\end{tabular} 
\\
\\
~\\
For example at $q=3$  the allowed ranges and saddles are,\\
~\\

\label{q3}
\begin{tabular} { | c | c | c | c | c |  c | } 
\hline
n = 0  &   n= 1 & n = 2 & n=3 & allowed $m^2$ range & saddle value of $m^2$ \\ 
\hline 
S & S & S & S &  $[0,\infty)$ & no solutions  \\
\hline
A & S & S & S & $[0,1]$ &  $\sim 0.071$ \\
\hline
A & A & S & S & $[0,\frac{8}{9}]$ &  $\sim 0.204$  \\
\hline
S & A & S & S & $[0,\frac{8}{9}]$ & no solutions \\
\hline
A & S & A & S & $[0,\frac{5}{9}]$ & $\sim 0.136$\\
\hline
S & S & A & S & $[0,\frac{5}{9}]$ &  no solutions     \\
\hline
A & A & A & S & $[0,\frac{5}{9}]$ &  $\sim 0.014, \sim 0.208$    \\
\hline
S & A & A & S & $[0,\frac{5}{9}]$ & $\sim 0.028$ \\
\hline 
\end{tabular} 
\\
\\
~\\
Now we need a prescription for regulating the value of $F_q$ so that it can be evaluated on these saddles for any value of $q$ and thus determine which is the global minima of $F_q$ and thus be able to determine the true state of the theory. 

\newpage


\section {Can we compare the free-energy $F_q$ at different saddles?} 

Earlier we had the following expression, 

\begin{eqnarray} 
\frac { 2F_q }{ N Vol( \mathbb{H}^2 ) } = \sum_{ n \in \mathbb{Z} } \int_{\mu >0}  d\mu \frac{ \tanh (\pi \sqrt{\mu} ) }{4 \pi  } \log ( \mu + m^2 + \frac{n^2}{q^2}  ) - \frac{2\pi q }{g_c}  (m^2 - \frac{1}{4}  )
\end{eqnarray}

On this we can use the zeta-function regularized identity, 

\begin{eqnarray*}
\sum_{n \in \mathbb{Z}} \log ( a^2 + \frac{ n^2}{q^2} ) = 2 \log ( 2 \sinh ( \pi q \vert a \vert ) )
\end{eqnarray*}

to write it as, 

\begin{eqnarray} 
\notag \frac { 2F_q }{ N Vol( \mathbb{H}^2 ) } =   \frac {1}{2\pi}   \int_{\mu >0}  d\mu  \tanh (\pi \sqrt{\mu} ) [   \pi q \sqrt{ \mu + m^2  }  + \log ( 1 - e^{ -2 \pi q \sqrt { \mu + m^2 }   }   )   ]  - \frac{2\pi q }{g_c}  (m^2 - \frac{1}{4}  )\\
\end{eqnarray}

The integral in the above expression is divergent and the divergence is not going to disappear by an analogous subtraction that was done previously. So for the moment we don't spend efforts into trying to regularize it but try to understand better the contribution of any specific ($n^{th}$) KK mode to $\frac { 2F_q }{ N Vol( \mathbb{H}^2 ) }$ and that is the integral, $ \int_{\lambda >0}  d\lambda \frac{ tanh (\pi \lambda ) }{2 \pi  } \log ( \lambda^2 + m^2 + \frac{n^2}{q^2}  ) $. (where we have substitute $\mu = \lambda^2$). Let $\vec{n}$ denote the tuple of integers corresponding to the KK modes being put into alternate quantization. For each $n_i \in \vec{n}$ we symbolically subtract the contribution of that mode in the standard quantization and add in its contribution in the alternate quantization. So we have,  

\begin{eqnarray*} 
\notag \frac { 2F_q^{(\vec{n},alt)}  }{ N Vol( \mathbb{H}^2 ) } &=   \frac {1}{2\pi}   \int_{\mu >0}  d\mu  \tanh (\pi \sqrt{\mu} ) [   \pi q \sqrt{ \mu + m^2  }  + \ln ( 1 - e^{ -2 \pi q \sqrt { \mu + m^2 }   }   )   ]  - \frac{2\pi q }{g_c}  (m^2 - \frac{1}{4}  )\\
&+ \sum_{n_i \in \vec{n}} \lim_{\Lambda \rightarrow \infty } \left [ [ - \int_{\lambda =0,standard} ^{\Lambda } + \int_{\lambda=0,alternate}^{\Lambda}]  d\lambda \frac{ \lambda \tanh (\pi \lambda ) }{2 \pi  } \log ( \lambda^2 + m^2 + \frac{n_i^2}{q^2}  ) \right ] 
\end{eqnarray*}
\newline 
In the above expression it is understood that the $m$ featuring on the right is the value of the saddle point mass (as calculated for example in the last section for $q=2,3$) if it exists for this combination of $\vec{n}$ being put in alternate quantization.  \emph {In the following we shall show that it is possible to give a finite meaning to the $\sum_{n_i \in \vec{n}}$ term} 

\newpage 

Now we define the two integrals $\int_{\lambda =0,standard} ^{\Lambda }$ and $ \int_{\lambda=0,alternate}^{\Lambda}$ as follows, \\ 
~\\ 

\begin{itemize}
\item We notice that the integrand is symmetric in changing $\lambda$ to $-\lambda$ and hence hence we can extend the integral to the line segment $[-\Lambda,\Lambda]$ and divide the final answer by $\frac{1}{2}$. Now we further imagine the integral to be in a complex $\lambda$ plane so that we are after going from $-\Lambda$ to $\Lambda$ we move up on a semicircle of radius $\Lambda$ centered at the origin till we get close to the $y-$axis. \\
~\\
Now we note that the function $\log(\lambda^2+m^2 + \frac{n_i^2}{q^2})$ needs two branch-cuts with one end at $\pm i\sqrt{m^2 + \frac{n_i^2}{q^2}}$. (\emph {this branch-cut has been explained in details in the Appendix}) We let the two branch-cuts start at $\pm i \sqrt{m^2 + \frac{n_i^2}{q^2}}$ and move up/down the $y-$axis. We also note that the function $tanh(\pi \lambda)$ has poles all along the $y-$axis at points $i(n+\frac{1}{2})$ for $n \in \mathbb{Z}$.\\    
~\\
Hence once the contour reaches the $y-$axis we drop down along its right side keeping a distance $\delta >0$ from the $y-$axis and making $\epsilon>0$ sized semi-circular humps centered around every $\tanh(\pi z)$ pole encountered till we are within $\epsilon$ of the branch-point $i\sqrt{m^2 + \frac{n_i^2}{q^2} }$. Then we make a clockwise circular tour around the point $i\sqrt{m^2 + \frac{n_i^2}{q^2} }$ in an $\epsilon$ radius circle and then move up the left side of the $y-$axis keeping a distance $\delta$ from it and as before making semi-circular humps around the $\tanh(\pi z)$ poles. Once we are at $-\delta + i\Lambda$ we go on to complete the semicircle of radius $\Lambda$ around the origin that we started out making.  It is understood that $\epsilon, \delta \rightarrow 0$.  \\
~\\
Hence we have defined $\int_{\lambda =0,standard}^{\Lambda} = \frac{1}{2}[-\int_{\Lambda,semicircle} - \int_{standard\text{ }key-hole}]$ with the meanings of the notation being obvious.  

\item Now when we try to define the quantity $\int_{\lambda=0,alternate}^{\Lambda}$, we do exactly as above except that instead of turning around and up around the branch-point $i\sqrt{m^2 + \frac{n_i^2}{q^2}}$ we keep going down into the lower half plane till we are $\epsilon$ \emph{away} from the other branch-point $-i\sqrt{m^2 + \frac{n_i^2}{q^2}}$ . When we are at $z = \delta -i(\sqrt{m^2 + \frac{n_i^2}{q^2} } + \epsilon)$, we start a \emph{counter-clockwise} tour around the negative imaginary branch-point and reach the point $-\delta -i(\sqrt{m^2 + \frac{n_i^2}{q^2} } + \epsilon) $ along this partial circle. It is clear that we take this circle so as to avoid intersecting the branch-cut which moves down the $y-$axis starting at the point $-i\sqrt{m^2 + \frac{n_i^2}{q^2}}$.\\
~\\
Then we continue moving up the $y-$ axis keeping a distance $\delta$ from it and making semi-circular $\epsilon$ radius humps around the $\tanh(\pi z)$ poles (as we also made on the right side of the $y-$axis while coming down). We move up till we are at $-\delta+i\Lambda$ and then we go on to complete the semicircle of radius $\Lambda$ around the origin that we started out making. \\
~\\
Hence we have defined $\int_{\lambda =0,alternate}^{\Lambda} = \frac{1}{2}[-\int_{\Lambda,semicircle} - \int_{alternate\text{ }key-hole}]$ with the meanings of the notation being obvious.  (we note that the part $\int_{\Lambda,semicircle}$ is exactly the same in both the quantization prescriptions)  
\end{itemize}

So the expression currently looks like,

\begin{eqnarray*} 
\notag \frac { 2F_q^{(\vec{n},alt)}  }{ N Vol( \mathbb{H}^2 ) }& = \frac {1}{2\pi}   \int_{\mu >0}  d\mu  \tanh (\pi \sqrt{\mu} ) [   \pi q \sqrt{ \mu + m^2  }  + log ( 1 - e^{ -2 \pi q \sqrt { \mu + m^2 }   }   )   ]  - \frac{2\pi q }{g_c}  (m^2 - \frac{1}{4}  )\\
\notag &+ \sum_{n_i \in \vec{n}} \lim_{\Lambda \rightarrow \infty } \left [ [ \frac{1}{2}[\int_{\Lambda,semicircle} + \int_{standard\text{ }key-hole}] d\lambda \frac{ \lambda \tanh (\pi \lambda ) }{2 \pi  } \log ( \lambda^2 + m^2 + \frac{n_i^2}{q^2}  ) \right ]  \\
\notag &+ \sum_{n_i \in \vec{n}} \lim_{\Lambda \rightarrow \infty } \left [ \frac{1}{2}[-\int_{\Lambda,semicircle} - \int_{alternate\text{ }key-hole}] ]  d\lambda \frac{ \lambda \tanh (\pi \lambda ) }{2 \pi  } \log ( \lambda^2 + m^2 + \frac{n_i^2}{q^2}  ) \right ] \\
\notag &= \frac {1}{2\pi}   \int_{\mu >0}  d\mu  \tanh (\pi \sqrt{\mu} ) [   \pi q \sqrt{ \mu + m^2  }  + log ( 1 - e^{ -2 \pi q \sqrt { \mu + m^2 }   }   )   ]  - \frac{2\pi q }{g_c}  (m^2 - \frac{1}{4}  )\\
\notag &+ \sum_{n_i \in \vec{n}}  \left [ \frac{1}{4\pi}[ \int_{standard\text{ }key-hole} - \int_{alternate\text{ }key-hole} ] dz \text{ }z \tanh (\pi z )  \log ( z^2 + m^2 + \frac{n_i^2}{q^2}  ) \right ]  \\
\end{eqnarray*}

Hence effectively the remnant quantity is  $\int_{standard\text{ }key-hole} - \int_{alternate\text{ }key-hole}$ and in it the contributions from the contours above the point $i(\epsilon + \sqrt{m^2 + \frac{n_i^2}{q^2} } )$ completely cancel out. What remains are these two parts,  

\begin{itemize}
\item There is a contour integral of the function $\lambda \tanh (\pi \lambda )  \log ( \lambda^2 + m^2 + \frac{n_i^2}{q^2})$ along a circular contour along an $\epsilon$ circle around the branch-points $\pm i \sqrt{m^2 + \frac{n_i^2}{q^2} }$ but avoiding intersecting the branch-cut at both of these.\\ 
~\\
One can imagine both these integrals to be of a holomorphic function on a Riemann surface such that the integration path doesn't close to a loop on the branch-cut. In that imagination the integration measure $``dz"$ scales with $\epsilon$ and hence in the limit these two integrations do not contribute. 

\item (From here on for the sake of ease of notation we denote $a_i = \sqrt{m^2 + \frac{n_i^2}{q^2} } $)\\ 
~\\
The integration path that does contribute is the journey from $-i(a_i+\epsilon)$ to $i(a_i+\epsilon)$ upwards and downwards along the right and the left of the $y-$axis keeping a distance of $\delta$ from it and making semi-circular humps of radius $\epsilon$ around the points $i(n + \frac{1}{2})$ ($n \in \mathbb{Z})$ that one encounters along the way. Let $i(n_{iu} + \frac{1}{2})$ be such a pole just below $ia_i$ and let $i(n_{id} + \frac{1}{2})$ be such a pole just above $-ia_i$. One notes that if $\{ a_i \} > \frac{1}{2}$ then $n_{id} = -(1+\lfloor a_i \rfloor ), n_{iu} = \lfloor a_i \rfloor$. If $\{ a_i \} \leq \frac{1}{2}$ then $n_{id} = - \lfloor a_i \rfloor, n_{iu} = \lfloor a_i \rfloor -1$.

\newpage 

We split this above integral into $3$ parts, 
\begin{enumerate} 

\item First we consider the neck that exists from $i(a_i+\epsilon)$ to $i\epsilon + i(n_{iu} +\frac{1}{2} )$. We parameterize the upward journey on the right of the $y-$axis as $z_R = i(a_i+\epsilon) + \delta -it$ from $t=a_i-(n_{iu} + \frac{1}{2} )$  to $t=0$ and the downward journey on the left as $z_L = i(a_i+\epsilon)-\delta -it$ for the same range of $t$ in the reverse order. (we remember that the direction of travel is upward on the right and downward on the left because of the ``-" sign infront of the $\int_{alternate\text{ }key-hole}$ integral) We note that an extra $2i\pi$ needs to be added to the $\log$ function when it is on the left of the $y-$axis and above the branch-point i.e on the $z_L$ path from $t=0$ to $t=\epsilon$. (\emph { this factor of $2i\pi$ has been derived in details in the Appendix}) So the corresponding integral is, 

\begin{eqnarray*}
\notag &\int_{t= a_i -(n_{iu} + \frac{1}{2} ) }^0 z_R \tanh(\pi z_R) \log(z_R^2 + a_i^2) dz_R +  \int_{t=0} ^{a_i -(n_{iu} + \frac{1}{2} ) } z_L \tanh(\pi z_L) \log(z_L^2 + a_i^2) dz_L\\
\notag &+ \int_0^\epsilon z_L \tanh (\pi z_L) [i2\pi ]dz_L \\
&(= 0) 
\end{eqnarray*} 

If we let $\epsilon, \delta \rightarrow 0$ then the integrands of the first two integrals are the same and hence the integrals cancel out because of their limits being in reverse order to each other. The last integral vanishes because in this limit its integration limits become the same.  

\item Now we consider the part of the contour that goes around the pole $i(n_{id}+\frac{1}{2})$ in anti-clockwise direction in a circle of radius $\epsilon$ and moves up/down the $y-$axis till $-i(a_i+\epsilon)$. Let the trip around the $n_{id}$ pole be parameterized as $z_c = i(n_{id} + \frac{1}{2})+\epsilon e^{i\phi}$ and on the left below it as $z_L = i(n_{id} + \frac{1}{2}) -i\epsilon - \delta -it$ and on the right as $z_R = i(n_{id} + \frac{1}{2})-i\epsilon + \delta -it$. For $z_L$ its parameter $t$ goes from $0$ to $a_i+(n_{id} + \frac{1}{2})$ and for $z_R$ its the same range of $t$ but in reverse order. Now we need to add an extra $2i\pi$ to $\log$ function in the $z_L$ integral when it is below the branch-point i.e for $t=a_i+(n_{id}+\frac{1}{2})-\epsilon$ to $t=a_i+(n_{id} + \frac{1}{2})$.  (\emph {this factor of $2i\pi$ has been derived in details in the Appendix}) So the corresponding integral is, 

\begin{eqnarray*}
\notag &\int_{t= a_i +(n_{id} + \frac{1}{2} ) }^0 z_R \tanh(\pi z_R) \log(z_R^2 + a_i^2) dz_R +  \int_{t=0} ^{a_i +(n_{id} + \frac{1}{2} ) } z_L \tanh(\pi z_L) \log(z_L^2 + a_i^2) dz_L\\
\notag &+ \int_{a_i + (n_{id} + \frac{1}{2} ) - \epsilon }^{a_i + (n_{id} + \frac{1}{2} ) } z_L \tanh (\pi z_L) [i2\pi ]dz_L + \int _{\phi = -\pi} ^{\pi} z_c \tanh (\pi z_c) \log (z_c^2 + a_i^2) dz_c \\
&(=  \int _{\phi = -\pi} ^{\pi} z_c \tanh (\pi z_c) \log (z_c^2 + a_i^2) dz_c )
\end{eqnarray*} gf

If we let $\epsilon, \delta \rightarrow 0$ then the integrands of the first two integrals are the same but their integration limits are in the reverse order and hence they cancel out in the limit. The third integral vanishes in this limit as its integration limits coalesce. The only contribution that survives is the last integral. 

\newpage 

\item The last structure that needs to be accounted for is the journey from the pole $i(n_{iu}+\frac{1}{2})$ to $i(\epsilon + n_{id} + \frac{1}{2})$. This journey has a repeating structure as in for each $i(k + \frac{1}{2})$ pole encountered we are going around it anti-clockwise in a circle of radius $\epsilon$ and then moving down/up on the left/right of the $y-$axis (keeping $\delta$ away from it) below that circle till we hit the top of the next $\epsilon$ circle below it i.e the $\epsilon$ circle around $i(k-1+\frac{1}{2})$. This goes on from $k= n_{iu}$ down to $k=n_{id} + 1$.\\ 
~\\
We parameterize the trip around the pole $i(k+\frac{1}{2})$ as $z_c = i(k+\frac{1}{2}) + \epsilon e^{i\phi}$. The trip on the right and below it can be given as $z_R = i(k +\frac{1}{2})-i\epsilon + \delta - it$. The trip on the left and below the $z_c$ circle is parameterized as $z_L = i(k +\frac{1}{2})-i\epsilon - \delta - it$. For $z_R$, $t$ goes from $1-2\epsilon$ to $0$ and for $z_L$ it goes through the same values but in reverse order. So the net contribution of this part of the contour is, 

\begin{eqnarray*} 
\notag &\sum_{k = n_{id} +1 }^{n_{iu}} \left [ \int_{t=0}^{1-2\epsilon} z_L \tanh(\pi z_L)\log (z_L^2 + a_i^2)dz_L +   \int_{t=1-2\epsilon}^{0} z_R \tanh(\pi z_R)\log (z_R^2 + a_i^2) dz_R  \right ] \\
\notag &+ \sum_{k = n_{id} +1 }^{n_{iu}} \left [\int_{\phi = -\pi}^{\pi} z_c \tanh (\pi z_c) \log (z_c^2 +a_i^2)dz_c \right ] \\
&(= \sum_{k = n_{id} +1 }^{n_{iu}} \left [\int_{\phi = -\pi}^{\pi} z_c \tanh (\pi z_c) \log (z_c^2 +a_i^2)dz_c \right ])
\end{eqnarray*}

Here too in the limit $\epsilon, \delta \rightarrow 0$ the integrands of first two integrals become the same but since their integration limits go in opposite directions they cancel out. Hence the only contribution that remains is just the third integral. 
\end{enumerate}
\end{itemize}
~\\
So combining the contributions from all the above $1+3$ parts of the contour we get the final result as, 

\begin{eqnarray} 
\notag \frac { 2F_q^{(\vec{n},alt)}  }{ N Vol( \mathbb{H}^2 ) } &= \frac {1}{2\pi}   \int_{\mu >0}  d\mu  \tanh (\pi \sqrt{\mu} ) [   \pi q \sqrt{ \mu + m^2  }  + \log ( 1 - e^{ -2 \pi q \sqrt { \mu + m^2 }   }   )   ]  - \frac{2\pi q }{g_c}  (m^2 - \frac{1}{4}  )\\
\notag &+ \sum_{n_i \in \vec{n}}  \left [ \frac{1}{4\pi}  \sum_{k = n_{id} }^{n_{iu}} [\int_{\phi = -\pi}^{\pi}dz_c \text{ }z_c \tanh (\pi z_c )  \log ( z_c^2 + m^2 + \frac{n_i^2}{q^2}  ) ] \right ]  \\
\end{eqnarray}

These $\phi$ integrals can be given by the residue theorem as, 

\begin{eqnarray} 
\notag &\int_{\phi = -\pi} ^{\pi} z_c \tanh (\pi z_c) \log(z_c^2+a_i^2) = 2\pi i Res_{i(k+\frac{1}{2})} [z_c \tanh (\pi z_c) \log (z_c^2 + a_i^2) ] \\
\notag &= 2\pi i (i(k+\frac{1}{2})) \frac{1}{\pi} \log [a_i^2 - (k+\frac{1}{2})^2]\\ 
&= -(2k+1)\log [a_i^2 - (k+\frac{1}{2})^2] 
\end{eqnarray}

\newpage 

Hence the simplified final expression is, 

\begin{eqnarray} \label {final} 
\notag \frac { 2F_q^{(\vec{n},alt)}  }{ N Vol( \mathbb{H}^2 ) } &= \frac {1}{2\pi}   \int_{\mu >0}  d\mu  \tanh (\pi \sqrt{\mu} ) [   \pi q \sqrt{ \mu + m^2  }  + \log ( 1 - e^{ -2 \pi q \sqrt { \mu + m^2 }   }   )   ]  - \frac{2\pi q }{g_c}  (m^2 - \frac{1}{4}  )\\
\notag &- \sum_{n_i \in \vec{n}}  \left [ \frac{1}{4\pi}  \sum_{k = n_{id} }^{n_{iu}} (1+2k)\log [ (m^2 + \frac{n_i^2}{q^2} ) - (k +\frac{1}{2})^2 ] \right ]  \\
\end{eqnarray}

We remind ourselves that for each $i$ index we had defined $a_i = \sqrt{m^2 + \frac{n_i^2}{q^2} }$. If $\{ a_i \} > \frac{1}{2}$ then $n_{id} = -(1+\lfloor a_i \rfloor ), n_{iu} = \lfloor a_i \rfloor$. If $\{ a_i \} \leq \frac{1}{2}$ then $n_{id} = - \lfloor a_i \rfloor, n_{iu} = \lfloor a_i \rfloor -1$

We note that this $\sum_{n_i \in \vec{n}}$ exists only if for the given $q$ and $\vec{n}$ (and hence the saddle point $m^2$) is such that the value of $\sqrt{m^2 + \frac{n_i^2}{q^2}}$ is large enough such that there exists a non-zero number of $i(k+\frac{1}{2})$ poles (of $\tanh(\pi z)$) on the $y-$axis between $\pm i\sqrt{m^2 + \frac{n_i^2}{q^2}}$. It turns out that for the $q=2$ and $q=3$ cases explicitly computed in the previous section this is not the case and hence there is no $\sum_{n_i \in \vec{n}}$ term for them!\\ 
~\\
Hence for all the saddles detected at $q=2$ and $q=3$ the corresponding \ref{final} equation has only the first two terms, the divergent $\mu$-integral and the $g_c$ dependent term. The divergent $\mu$-integral has its integrand dependent on $q$ and $\vec{n}$ and hence it is tempting to redefine L.H.S of \ref{final} to absorb the divergent integral and compare the remaining finite terms to determine the physical minima. Doing this in the particular case of $q=2$ and $q=3$  leads to the conclusion that the q-Free-energy density is minimum for whoever has the minimum $g_c$ term and that is true for whichever saddle has the heaviest mass and hence that is the true state of the theory. 

\section {Conclusion}
A particularly novel interpretation of the meaning of ``alternate quantization'' has been explored in this writing in the specific context of Renyi entropy of a critical $O(N)$. It seems exciting to further explore the ramifications of this interpretation in other scenarios and to hopefully uncover a possibly more general definition of ``alternate quantization''. Immediately it seems interesting to try to generalize this calculation to other more general sigma models. Even in the context of the specific question explored in this paper, we are faced with a whole range of unanswered questions which we shall briefly enunciate here. Firstly it should be interesting to understand better as to how the formalism of this article is related to the analysis in section $6$ of the pioneering work \cite{Max} by Max A. Metlitski, Carlos A. Fuertes and Subir Sachdev. Secondly, as of now no structure is visible in the table of saddle points, \ref{q2} and \ref{q3}. One would ideally want to have an analytic method of knowing as to which combinations of KK modes being put into alternate quantization produces saddle points and where. This seems like a major question that is currently open here. Thirdly, the interpretation of \ref{final} is currently not on a very firm ground. One would hope to get a cleaner interpretation about how this equation helps us determine the true state of the theory. The theory of Renyi entropy has clearly opened up a new and exciting avenue of research in holography and this article hopefully shows how this can provide us a context to explore this otherwise rare question of understanding ground states of non-supersymmetric QFTs.      

\section {Acknowledgement}
The author is extremely grateful to Thomas Faulkner for the extensive discussions which helped formulate all the crucial techniques in this work. Without the enormous help of Thomas Faulkner this work would have never happened. The author was supported by the University of Illinois at Urbana-Champaign (UIUC) during the writing of the first half of this work. A lot of gratitude is due to Max Metlitski for providing a lot of insights and inspiration during the later stages of this work. 

\appendix 

\section {The branch-cut of the logarithm }

The logarithm function can be thought to have a branch-cut along the negative $x-$ axis which mathematically can be stated as, $lim_{\theta \rightarrow \pm \pi} Im [ log (r e^{i \theta } ) ]  = \pm \pi$ (though $\theta = \pm \pi$ is geometrically the same location). Now we want to do this same analysis for the function $log(z^2 + a^2)$. 

\subsection { $\log(z^2+a^2)$ about $ia$} 

Consider the function $\log (z^2 + a^2)$ while taking $z$ around in a small circle of radius $r < a $ about $ia$ for $a>0$. We parameterize this roundtrip  as $z = ia + re^{i\phi}$. Hence we are looking at the function $f(\phi) = log ((ia + r e^{i\phi})^2 + a^2 )$. This when expanded out becomes,  $f(\phi) = log ( (r^2 \cos 2 \phi  - 2 ar \sin \phi  ) + i (r^2 \sin  2\phi + 2ar \cos  \phi ) )$\\ 
~\\
Let  $\phi = \frac{\pi}{2} + \epsilon$, i.e when this trip is just crossing the positive $y-$axis. Then the real part of the argument of the logarithm evaluates to $-r^2\cos 2 \epsilon - 2 a r \cos \epsilon$. This for $\epsilon \rightarrow 0^{\pm}$ is a negative number. Hence the logarithm is being evaluated around its branch-cut which is the negative $x-$axis.\\ 
~\\ 
At the same location the imaginary part of the argument of the logarithm function evaluates to $-r^2 \sin 2 \epsilon + 2 a r \sin \epsilon$ and this for $\epsilon \rightarrow 0^{\pm}$ is $2r \epsilon (a- r )$. Since $r < a$ it follows that for $\epsilon>0$ the log function is being evaluated just above its negative $x-$axis branch-cut and for $\epsilon < 0$ it is just below that.\\ 
~\\
So because of how the phase of the complex logarithm function changes we have that, $f(\frac{\pi}{2}^{+}) = 2i\pi + f(\frac{\pi}{2}^{-})$. This gives meaning to the statement that one of the branch-cuts of the function $f(z) = log(z^2+a^2)$ starts at $ia$ and goes up the positive $y-$axis. (so a $2i\pi$ has to be added to the log function on the left of this branch-cut)

\subsection {$\log(z^2+a^2)$ about $-ia$}

Consider the function $log (z^2 + a^2)$ while taking $z$ around in a small circle of radius $r < a $ about $-ia$ for $a>0$. We parametrize this roundtrip  as $z = -ia + re^{i\phi}$. Hence we are looking at the function $f(\phi) = log ((-ia + r e^{i\phi})^2 + a^2 )$. This when expanded out becomes,  $f(\phi) = log ( (r^2 \cos 2 \phi  + 2 ar \sin \phi  ) + i (r^2 \sin  2\phi - 2ar \cos  \phi ) )$\\ 
~\\
Let  $\phi = -\frac{\pi}{2} + \epsilon$, i.e when this trip is just crossing the negative $y-$axis. Then the real part of the argument of the logarithm evaluates to $-r^2\cos 2 \epsilon - 2 a r \cos \epsilon$. This for $\epsilon \rightarrow 0^{\pm}$ is a negative number. Hence the logarithm is being evaluated around its branch-cut which is the negative $x-$axis.\\ 
~\\ 
At the same location the imaginary part of the argument of the logarithm function evaluates to $-r^2 \sin 2 \epsilon - 2 a r \sin \epsilon$ and this for $\epsilon \rightarrow 0^{\pm}$ is $-2r \epsilon (a +r )$. Hence it follows that for $\epsilon>0$ the log function is being evaluated just below its negative $x-$axis branch-cut and for $\epsilon < 0$ it is just above that.\\ 
~\\
So because of how the phase of the complex logarithm function changes we have that, $f(-\frac{\pi}{2}^{-}) = 2i\pi + f(-\frac{\pi}{2}^{+})$. This gives meaning to the statement that one of the branch-cuts of the function $f(z) = log(z^2+a^2)$ starts at $-ia$ and goes down the negative $y-$axis. (so a $2i\pi$ has to be added to the log function on the left of this branch-cut)

\end{document}